\documentclass[twocolumn]{revtex4}
\usepackage{graphicx}
\usepackage{epsf}
\def\be{\begin{equation}}
\def\ee{\end{equation}}
\def\ba{\begin{array}}
\def\ea{\end{array}}
\def\bea{\begin{eqnarray}}
\def\eea{\end{eqnarray}}
\def\drm{{\mathrm d}}
\def\erm{{\mathrm e}}
\def\dps{\displaystyle}
\def\rt{\rightarrow}
\def\ul{\underline}
\def\n{\nu}
\def\t{\tau}
\begin{document}

\vspace{-6truecm} %
\noindent DSF$-$02/2006 %
%physics/0603140 %
\vspace{1truecm}

\title{Majorana and the path-integral approach to Quantum Mechanics}%
\author{S. Esposito}%
\address{{\it S. Esposito}: Dipartimento di Scienze Fisiche,
Universit\`a di Napoli ``Federico II'' \& I.N.F.N. Sezione di
Napoli, Complesso Universitario di M. S. Angelo, Via Cinthia,
80126 Napoli ({\rm Salvatore.Esposito@na.infn.it})}%

%\thanks{}%
%\subjclass{}%
%\keywords{}%

%\date{}%
%\dedicatory{}%
%\commby{}%
%----------------------------------------------------------------
\begin{abstract}
We give, for the first time, the English translation of a
manuscript by Ettore Majorana, which probably corresponds to the
text for a seminar delivered at the University of Naples in 1938,
where he lectured on Theoretical Physics. Some passages reveal a
physical interpretation of the Quantum Mechanics which anticipates
of several years the Feynman approach in terms of path integrals,
independently of the underlying mathematical formulation.
\end{abstract}

\maketitle

%----------------------------------------------------------------
\section{Introduction}

\noindent The interest in the course on Theoretical Physics,
delivered by Ettore Majorana at the University of Naples in 1938,
has been recently revived by the discovery of the Moreno Paper
\cite{moreno}, which is a faithful transcription of the lecture
notes prepared by Majorana himself made by the student Eugenio
Moreno. Such a Paper, in fact, includes some previously unknown
lecture notes, whose original manuscripts seem to be missing. The
handwritten notes by Majorana are now kept at the Domus Galilaeana
in Pisa, and were anastatically reproduced some years ago
\cite{Bibliopolis} by including also some papers, initially
interpreted as the notes prepared for a forthcoming lecture that,
however, Majorana never gave due to his mysterious disappearance.

As recently shown \cite{espoquad}, instead, such spare papers
cannot be considered as notes for academic lectures, even for an
advanced course as that delivered by Majorana, but probably refer
to a general conference given by him at a restricted audience
interested in Molecular Physics. Although the main topic of that
dissertation was the application of Quantum Mechanics to the
theory of molecular bonding, the present scientific interest in it
is more centered on the interpretation given by Majorana about
some topics of the novel, for that time, Quantum Theory (namely,
the concept of quantum state) and the direct application of this
theory to a particular case (that is, precisely, the molecular
bonding). An accurate reading of the manuscript, in fact, not only
discloses a peculiar cleverness of the author in treating a
pivotal argument of the novel Mechanics, but, keeping in mind that
it was written in 1938, also reveals a net advance of at least ten
years in the use made of that topic. The last point, however, is
quite common for Majorana, and as only one example we refer here
to the case of the Thomas-Fermi atomic model \cite{fermithomas}
(see, however, also Ref. \cite{volumetti}).

This point was already noted some years ago by N. Cabibbo (in Ref.
\cite{Bibliopolis}), who saw in the Majorana manuscript a vague
and approximate anticipation of the idea underlying the Feynman
interpretation of Quantum Mechanics in terms of path integrals. A
more analytic study, conducted on the critical edition of that
paper \cite{espobiblio} which was not available at that time, but
is here reported, for the first time, in translation, reveals
instead some intriguing surprises, upon which we will here focus
on. Although the text considered is written in a very simple a
clear form (a feature which is very common in the Majorana works
\cite{volumetti}), in the next section we give beforehand a brief
discussion of the main ideas of the path-integral approach,
followed by a simple presentation where the crucial passages in
the Majorana paper are pointed out.

\section{Quantum Mechanics in the path-integral approach}

\noindent The general postulate upon which Quantum Mechanics is
based tells that the ``state" of a certain physical system may be
represented with a complex quantity $\psi$, considered as a
(normalized) vector in a given Hilbert space corresponding to the
physical system, where all the information on the system is
contained \cite{caldirola}. The time evolution of the state vector
is ruled by the Schr\"odinger equation, that may be written in the
general form: %
\be \label{e1} %
i \hbar \frac{\drm \psi}{\drm t} = H \, \psi ,
\ee %
where $H$ is the hamiltonian operator of the considered system.
The initial state at time $t_0$ is specified by its choice among
the possible eigenstates of a complete set of commuting operators
(including, for example, the hamiltonian), while the hamiltonian
$H$ itself determines the state of the system at a subsequent time
$T$ through Eq. (\ref{e1}). The dynamical evolution of the system
is thus completely determined if we evaluate the transition
amplitude between the state at time $t_0$ and that at time $t$.

Then, as we can easily see, the usual quantum-mechanical
description of a given system is strongly centered on the role
played by the hamiltonian $H$ and, as a consequence, the time
variable plays itself a key role in this description.  Such a
dissymmetry between space and time variables is, obviously, not
satisfactory in the light of the postulates of the Theory of
Relativity. This was firstly realized in 1932 by Dirac
\cite{dirac}, who put forward the idea of reformulating the whole
Quantum Mechanics in terms of lagrangians rather than
hamiltonians\footnote{Note, however, that the Quantum Theory of
wave fields was already formulated in terms of a variational
principle applied to a lagrangian function. As a reference, see
the classical book by Heisenberg \cite{Heisenberg}, considered
also by Majorana for some works of him as, for example, the
development of a relativistic theory of particles with arbitrary
spin \cite{teorel} or the symmetric theory of electron and
positron \cite{elpos}.}

The starting point in the Dirac thought is that of exploiting an
analogy, holding at the quantum level, with the Hamilton principal
function in Classical Mechanics \cite{goldstein}. From this, the
transition amplitude from a state in the spatial configuration
$q_a$ at time $t_a$ to a state in the spatial configuration $q_b$
at time $t_b$ is written as:
\be \label{e2}%
\langle q_b | q_a \rangle \sim \erm^{\dps \frac{i}{\hbar} S} =
\erm^{\dps \frac{i}{\hbar} \int_{t_a}^{t_b} \!\! L \drm t}
\ee%
where $L$ is the lagrangian of the system and $S[q]$ the action
functional defined on the paths from $q_a$ to $q_b$. The previous
relationship, however, cannot be considered as an equality as long
as the time interval from $t_a$ to $t_b$ is finite, since it would
lead to incorrect results (and Dirac himself was well aware of
this; in his paper he introduced several unjustified assumptions
in order to overcome such a difficulty). In fact, by splitting the
integration field in (\ref{e2}) into $N$ intervals, $t_a = t_0 <
t_1 < t_2 < \dots < t_{N-1} < t_N = t_b$, the transition amplitude
could be written as a product of terms, %
\be \label{e3} %
\langle q_b | q_a \rangle = \langle q_b | q_{N-1} \rangle \langle
q_{N-1} | q_{N-2} \rangle \cdots \langle q_2 | q_1 \rangle \langle
q_1 | q_a \rangle ,
\ee %
while it is well known that, by using the completeness relations,
the correct formula contains the integration over the intermediate
regions: %
\bea %
& & \!\!\!\!\!\! \langle q_b | q_a \rangle = \langle q_b | \! \int
\!\! \drm q_{N-1} \, | q_{N-1} \rangle \langle q_{N-1} | \dots \!
\int
\!\! \drm q_{1} \, | q_{1} \rangle \langle q_{1} | q_a \rangle \nonumber \\
& & \!\! = \int \!\! \drm q_1 \drm q_2 \cdots \drm q_{N-1} \,
\langle q_b | q_{N-1} \rangle \cdots \langle q_2 | q_1 \rangle
\langle q_1 | q_a \rangle . \label{e4}
\eea %
About ten year after the appearance of the Dirac paper, Feynman
\cite{feynman} guessed that the relation in (\ref{e2}) should hold
as an equality, up to a constant factor $A$, only for transitions
between states spaced by an {\it infinitesimal} time interval. In
this case, by employing the correct formula in (\ref{e4}), we
obtain the well-known Feynman expression for the transition
amplitude between two given states: %
\bea  %
\langle q_b | q_a \rangle &=& \!\!\!\!\!\!\!\!\!\!\!\!\!\!\!\!\!\!
\lim_{\scriptsize ~~~~~~~~~~~ \ba{l} N \rt \infty \\
t_b - t_a \, {\rm finito} \ea} \!\!\!\! A^N \! \int \!\! \drm q_1
\drm q_2 \cdots \drm q_{N-1} \, \erm^{i S/\hbar} \nonumber  \\
& \equiv & \int \! {\rm D}q \, \erm^{i S/\hbar} . \label{e5}
\eea %
The meaning of the previous formula can be understand as follows.
In the integrations present in it, which we have generically
denoted with $\dps \int \! {\rm D}q$, the limits $t_a$ and $t_b$
in the integration interval are kept fixed, while we integrate
over the space specified by the intermediate points. Since every
spatial configuration $q_i$ of these intermediate points
corresponds to a given dynamical trajectory joining the initial
point $t_a$ with the final one $t_b$, the integration over all
these configurations is equivalent to sum over {\it all} the
possible paths from the initial point to the final one. In other
words the Feynman formula with the path integrals points out that
the transition amplitude between an initial and a final state can
be expressed as a sum of the factor $\erm^{iS[q]/\hbar}$ over all
the paths with fixed end-points. Such a result, on one hand, it is
not surprising if we consider that, in Quantum Mechanics, for a
given process taking place through different ways, the transition
amplitude is given by the sum of the partial amplitudes
corresponding to {\it all} the possible ways through which the
process happens. This is particularly evident, for example, in the
classical experiment with an electron beam impinging on a screen
through a double slit. The interference pattern observed on the
screen at a given distance from the slit, should suggest that a
single electron has crossed through both slits. It can be, then,
explained by assuming that the probability that has one electron
to go from the source to the screen through the double slit is
obtained by summing over {\it all} the possible paths covered by
the electron. Thus the fundamental principles of Quantum Mechanics
underlie the path-integral approach. However, what is crucial but
unexpected in this approach is that the sum is made over the phase
factor $\erm^{iS[q]/\hbar}$, which is generated by the classical
action $S[q]$.

One of the major merits of the Feynman approach to Quantum
Mechanics is the possibility to get in a very clear manner the
classical limit when $\hbar \rt 0$ (the other merits being its
versatile applicability to field quantization in abelian or non
abelian gauge theories, with or without spontaneous symmetry
breaking). In fact for large values of $S$ compared to $\hbar$,
the phase factor in (\ref{e5}) undergoes large fluctuations and
thus contributes with terms which average to zero. From a
mathematical point of view, it is then clear that in the limit
$\hbar \rt 0$ the dominant contribution to Eq. (\ref{e5}) comes
out when the phase factor does not varies much or, in other words,
when the action $S$ is stationary. This result is precisely what
emerges in Classical Mechanics, where the classical dynamical
trajectories are obtained from the least action principle. This
occurrence was firstly noted by Dirac \cite{dirac}, who realized
the key role played by the action functional.

The intuitive interpretation of the classical limit is very
simple. Let us consider, in the $q,t$ space, a given path far away
from the classical trajectory $q_{\rm{cl}}(t)$; since $\hbar$ is
very small, the phase $S/\hbar$ along this path will be quite
large. For each of such paths there is another one which is
infinitely close to it, where the action $S$ will change only for
a small quantity, but since this is multiplied by a very large
constant ($1/\hbar$), the resulting phase will have a
correspondingly large value. On average these paths will give a
vanishing contribution in the sum in Eq. (\ref{e5}). Instead, near
the classical trajectory $q_{\rm{cl}}(t)$ the action is
stationary, so that, passing by a path infinitely close to that
classical to this last one, the action will not change at all.
Then the corresponding contributions in Eq. (\ref{e5}) will sum
coherently and, as a result, the dominant term is obtained for
$\hbar \rt 0$. Thus, in such approach, the classical trajectory is
picked out in the limit $\hbar \rt 0$ not because it mainly
contributes to the dynamical evolution of the system, but rather
because there are paths infinitely close to it that give
contributions which sum coherently. The integration region is,
actually, very narrow for classical systems, while it becomes
wider for quantum ones. As a consequence, the concept of orbit
itself, that in the classical case is well defined, for quantum
systems comes to lose its meaning, as for example for an electron
orbiting around the atomic nucleus.

\section{Majorana contribution}

\noindent The discussion presented above of the path-integral
approach to Quantum Mechanics has been deliberately centered on
the mathematical aspect, rather than on the physical one, since
the historical process has precisely followed this line, starting
from the original Dirac idea in 1932 and arriving at the Feynman
formulation in the forties (one can usefully consult the original
papers in \cite{dirac} and \cite{feynman}). On the other hand,
just the development of the mathematical formalism has later led
to the impressive physical interpretation mentioned previously.

Coming back to the Majorana paper here considered, we immediately
realize that it contains {\it nothing} of the mathematical aspect
of the peculiar approach to Quantum Mechanics. Nevertheless an
accurate reading reveals, as well, the presence of the {\it
physical} foundations of it.

The starting point in Majorana is to search for a meaningful and
clear formulation of the concept of quantum state. And, obviously,
in 1938 the dispute is opened with the conceptions of the Old
Quantum Theory.
\begin{quote}
According to the Heisenberg theory, a quantum state corresponds
not to a strangely privileged solution of the classical equations
but rather to a set of solutions which differ for the initial
conditions and even for the energy, i.e. what it is meant as
precisely defined energy for the quantum state corresponds to a
sort of average over the infinite classical orbits belonging to
that state. Thus the quantum states come to be the minimal
statistical sets of classical motions, \ul{slightly different}
from each other, accessible to the observations. These minimal
statistical sets cannot be further partitioned due to the
uncertainty principle, introduced by Heisenberg himself, which
forbids the precise simultaneous measurement of the position and
the velocity of a particle, that is the determination of its
orbit.
\end{quote}
Let us note that the ``solutions which differ for the initial
conditions" correspond, in the Feynman language of 1948, precisely
to the different integration paths. In fact, the different initial
conditions are, in any case, always referred to the same initial
time ($t_a$), while the determined quantum state corresponds to a
fixed end time ($t_b$). The introduced issue of ``\ul{slightly
different} classical motions" (the emphasis is given by Majorana
himself), according to what specified by the Heisenberg's
uncertainty principle and mentioned just afterwards, is thus
evidently related to that of the sufficiently wide integration
region in Eq. (\ref{e5}) for quantum systems. In this respect,
such a mathematical point is intimately related to a fundamental
physical principle.

The crucial point in the Feynman formulation of Quantum Mechanics
is, as seen above, to consider not only the paths corresponding to
classical trajectories, but {\it all} the possible paths joining
the initial point with the end one. In the Majorana manuscript,
after a discussion on an interesting example on the harmonic
oscillator, the author points out:
\begin{quote}
Obviously the correspondence between quantum states and sets of
classical solutions is only approximate, since the equations
describing the quantum dynamics are in general independent of the
corresponding classical equations, but denote a real modification
of the mechanical laws, as well as a constraint on the feasibility
of a given observation; however it is better founded than the
representation of the quantum states in terms of quantized orbits,
and can be usefully employed in qualitative studies.
\end{quote}
And, in a later passage, it is more explicitly stated that the
wave function ``corresponds in Quantum Mechanics to any possible
state of the electron". Such a reference, that only superficially
could be interpreted, in the common acceptation, that all the
information on the physical systems is contained in the wave
function, should instead be considered in the meaning given by
Feynman, according to the comprehensive discussion made by
Majorana on the concept of state.

Finally we point out that, in the Majorana analysis, a key role is
played by the symmetry properties of the physical system.
\begin{quote}
Under given assumptions, that are verified in the very simple
problems which we will consider, we can say that every quantum
state possesses all the symmetry properties of the constraints of
the system.
\end{quote}
The relationship with the path-integral formulation is made as
follows. In discussing a given atomic system, Majorana points out
how from one quantum state $S$ of the system we can obtain another
one $S'$ by means of a symmetry operation.
\begin{quote}
However, differently from what happens in Classical Mechanics for
the \ul{single solutions} of the dynamical equations, in general
it is no longer true that $S'$ will be distinct from $S$. We can
realize this easily by representing $S'$ with a set of classical
solutions, as seen above; it then suffices that $S$ includes, for
any given solution, even the other one obtained from that solution
by applying a symmetry property of the motions of the systems, in
order that $S'$ results to be identical to $S$.
\end{quote}
This passage is particularly intriguing if we observe that the
issue of the redundant counting in the integration measure in
gauge theories, leading to infinite expressions for the transition
amplitudes \cite{ridondanza}, was raised (and solved) only after
much time from the Feynman paper.

Summing up, it is without doubt that no trace can be found of the
formalism underlying the Feynman path-integral approach to Quantum
Mechanics in the Majorana manuscript (on the contrary to what
happens for the Dirac paper of 1933, probably known to Majorana).
Nevertheless it is very interesting that the main physical items,
about the novel way of interpreting the theory of quanta, were
realized well in advance by Majorana. And this is particularly
impressive if we take into account that, in the known historical
path, the interpretation of the formalism has only followed the
mathematical development of the formalism itself.

Furthermore, in the Majorana paper, several interesting
applications to atomic and molecular systems are present as well,
where known results are deduced or re-interpreted according to the
novel point of view. The search for such applications, however,
will be left to the reader, who will benefit of the reading of the
complete text by Majorana reported in the following.

\section{The text by Majorana}

\noindent The original manuscript by Majorana, as it can be seen
from Ref. \cite{Bibliopolis}, reports (at the beginning) a sort of
table of contents which, however, is only partially followed by
the author. For the sake of ease, we have preferred to divide the
whole text in some sections, according to the reported table of
contents.

\setcounter{equation}{0}
\renewcommand{\theequation}{\Roman{equation}}

\renewcommand{\thesubsection}{\arabic{subsection}}

\subsection{On the meaning of quantum state}

\noindent The internal energy of a closed system (atom, molecule,
etc.) can take, according to Quantum Mechanics, discrete values
belonging to a set $E_0,E_1,E_2,\dots$ composed of the so-called
energy "eigenvalues". To each given value of the energy we can
associate a ``quantum state", that is a state where the system may
remain indefinitely without external perturbations. As an example
of these perturbations, we can in general consider the coupling of
the system with the radiation field, by means of which the system
may loose energy in form of electromagnetic radiation, jumping
from an energy level $E_k$ to a lower one $E_i<E_k$. Only when the
internal energy takes the minimum value $E_0$, it cannot be
further decrease by means of radiation; in this case the system is
said to be in its ``ground state" from which it cannot be removed
without sufficiently strong external influences, such as the
scattering with fast particles or with light quanta of large
frequency.
\\
What is the corresponding concept of quantum state in Classical
Mechanics? An answer is primarily required to such a question, in
order to have a correct representation of the results obtained in
our field by Quantum Mechanics, without entering, however, in the
complex computational methods adopted by this. \\
In Classical Mechanics the motion of a system composed of $N$ mass
points is entirely determined when the coordinates $q_1 \dots
q_{3N}$ of all the points are known as function of time: %
\be \label{a1} %
q_i = q_i(t)
\ee %
Eqs. (\ref{a1}) give the dynamical equations where all the
internal and external forces acting on the system are present, and
they can always be chosen in such a way that at a given instant
all the coordinates $q_i(0)$ and their time derivatives
$\dot{q}_i(0)$ take arbitrarily fixed values. Thus the general
solution of the equations of motion must depend $2 \cdot 3 N$
arbitrary constants. \\
For system with atomic dimensions the classical representation no
longer holds and two successive modifications have been proposed.
The first one, due to Bohr and Sommerfeld and that has provided
very useful results, has been afterwards completely abandoned with
the emerging of the novel Quantum Mechanics, which has been the
only one to give an extremely general formalism, fully confirmed
by the experiences on the study of the elementary processes.
According to the old theory of Bohr-Sommerfeld, Classical
Mechanics still holds in describing the atom, so that the motion
of an electron, for example, around the hydrogen nucleus is still
described by a solution (\ref{a1}) of the equations of Classical
Mechanics; however, if we consider periodic motions, such as the
revolution of an electron around the nucleus, not all the
solutions of the classical equations are realized in Nature, but
only a discrete infinity of those satisfying the so-called
Sommerfeld conditions, that is certain cabalistic-like integral
relations. For example in every periodic motion in one dimension,
the integral of the double of the kinetic energy over the period
$\t$:
\[ \int_0^\t \!\! 2 T(t) \, \drm t = n h \]
must be an integer multiple of the Planck constant ($h=6.55 \cdot
10^{-27}$). The combination of Classical Mechanics with a
principle which is unrelated with it, such as that of the
quantized orbits, appears so hybrid that it should be not
surprising the complete failure of that theory occurred in the
last decade, irrespective of several favorable experimental tests
which was supposed to be conclusive. \\
The novel Quantum Mechanics, primarily due to Heisenberg, is
substantially more closed to the classical conceptions than the
old one. According to the Heisenberg theory, a quantum state
corresponds not to a strangely privileged solution of the
classical equations but rather to a set of solutions which differ
for the initial conditions and even for the energy, i.e. what it
is meant as precisely defined energy for the quantum state
corresponds to a sort of average over the infinite classical
orbits belonging to that state. Thus the quantum states come to be
the minimal statistical sets of classical motions, \ul{slightly
different} from each other, accessible to the observations. These
minimal statistical sets cannot be further partitioned due to the
uncertainty principle, introduced by Heisenberg himself, which
forbids the precise simultaneous measurement of the position and
the velocity of a particle, that is the determination of its
orbit. \\
An harmonic oscillator with frequency $\n$ can oscillate
classically with arbitrary amplitude and phase, its energy being
given by %
\[ E = 2 \pi^2 m \n^2 A_0^2 \]
where $m$ is its mass and $A_0$ the maximum elongation. According
to Quantum Mechanics the possible values for $E$ are, as well
known, $\dps E_0 = \frac{1}{2} h \n$, $\dps E = \frac{3}{2} h \n$,
\dots $\dps E_n = \left( n + \frac{1}{2} \right) h \n$ \dots; in
this case we can say that the ground state with energy $\dps E_0 =
\frac{1}{2} h \n$ corresponds roughly to all the classical
oscillations with energy between $0$ and $h \n$, the first excited
state with energy $\dps E_0 = \frac{3}{2} h \n$ corresponds to the
classical solutions with energy between $h \n$ and $2 \cdot h \n$,
and so on. Obviously the correspondence between quantum states and
sets of classical solutions is only approximate, since the
equations describing the quantum dynamics are in general
independent of the corresponding classical equations, but denote a
real modification of the mechanical laws, as well as a constraint
on the feasibility of a given observation; however it is better
founded than the representation of the quantum states in terms of
quantized orbits, and can be usefully employed in qualitative
studies.

\subsection{Symmetry properties of a system in Classical and
Quantum Mechanics}

\noindent Systems showing some symmetry property deserve a
particular study. For these systems, due to symmetry
considerations alone, from one particular solution of the
classical equations of motion $q_i = q_i(t)$ we can deduce, in
general, some other different ones $q_i' = q_i'(t)$. For example
if the system contains two or more electrons or, in general, two
or more identical particles, from one given solution we can obtain
another solution, which in general will be different from the
previous one, just by changing the coordinates of two particles.
Analogously if we consider an electron moving in the field of two
identical nuclei or atoms (denoted with $A$ and $B$ in the
figure), starting from an allowed orbit $q_i = q_i(t)$ described
around $A$ with a given law of motion, we can deduce another orbit
$q_i' = q_i'(t)$ described by the electron around the nucleus or
atom $B$ by a reflection with respect to the center $O$ of the
line $AB$.
\begin{center}
%\begin{figure}[t]
\epsfysize=2cm \epsfxsize=6.5truecm %
\centerline{\epsffile{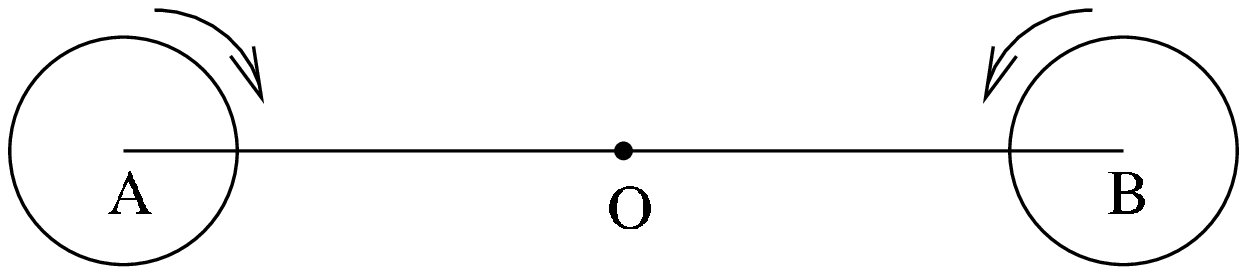}}
%\caption{}
%\label{app1fig}
%\end{figure}
\end{center} \vspace{-0.5truecm}
The exchange operations between two identical particles,
reflection with respect to one point or other ones corresponding
to any symmetry property, keep their meaning in Quantum Mechanics.
Thus it is possible to deduce from a state $S$ another one $S'$,
\ul{corresponding to the same known value of the energy}, if in
the two mentioned examples we exchange two identical particle
between them and reflect the system with respect to the point $O$.
However, differently from what happens in Classical Mechanics for
the \ul{single solutions} of the dynamical equations, in general
it is no longer true that $S'$ will be distinct from $S$. We can
realize this easily by representing $S'$ with a set of classical
solutions, as seen above; it then suffices that $S$ includes, for
any given solution, even the other one obtained from that solution
by applying a symmetry property of the motions of the systems, in
order that $S'$ results to be identical to $S$. \\
In several cases, if the system satisfies sufficiently complex
symmetry properties, it is instead possible to obtain, by symmetry
on a given quantum state, other different states but with the same
energy. In this case the system is said to be \ul{degenerate}, i.e
it has many states with the same energy, exactly due to its
symmetry properties. The study of degenerate systems and of the
conditions under which degeneration can take place will bring us
too far and, in any case, it is difficult to made such a study in
terms of only classical analogies. Then we will leave it
completely aside and limit our attention to problems without
degeneration. This condition is always satisfied if the symmetry
of the mechanical system allows only a so simple transformation
that its square, that is the transformation applied twice, reduces
to the identity transformation. For example, by a double
reflection of a system of mass points with respect a plane, a line
or a point, we necessarily recover the same initial arrangement;
analogously, the system remains unaltered by changing twice two
identical particles. In all these cases we have only simple
quantum states, i.e. to every possible value of the energy it is
associated only one quantum state. It follows that all the quantum
states of system containing two identical particles are  symmetric
with respect to the two particles, remaining unaltered under their
exchange. Thus the states of an electron orbiting around two
identical nuclei $A$ and $B$ are symmetric with respect to the
middle point $O$ of $AB$, or remain unaltered  by reflection in
$O$, and analogously for other similar cases. Under given
assumptions, that are verified in the very simple problems which
we will consider, we can say that every quantum state possesses
all the symmetry properties of the constraints of the system.

\subsection{Resonance forces between states that cannot be
symmetrized for small perturbations and spectroscopic
consequences. Theory of homeopolar valence according to the method
of bounding electrons. Properties of the symmetrized states that
are not obtained from non symmetrized ones with a weak
perturbation.}

\noindent Let us consider an electron moving in the field of two
hydrogen nuclei or protons. The system composed by the two protons
and the electron has a net resulting charge of $+e$ and
constitutes the simplest possible molecule, that is the positively
ionized hydrogen molecule. In such a system the protons are able
to move as well as the electron, but due to the large mass
difference between the first ones and the second one (mass ratio
1840:1) the mean velocity of the protons is much lower than that
of the electron, and the motion of this can be studied with great
accuracy by assuming that the protons are at rest at a given
mutual distance. This distance is determined, by stability
reasons, in such a way that the total energy of the molecule, that
at a first approximation is given by the sum of the mutual
potential energy of the two protons and the energy of the electron
moving in the field of the first ones, and is different for
different electron quantum states, is at a minimum. \\
The mutual potential energy of the protons is given by $\dps
\frac{e^2}{r}$ if $r$ is their distance, while the binding energy
of the electron in its ground state is a negative function $E(r)$
of $r$ that does not have a simple analytic expression, but it can
be obtained from Quantum Mechanics with an arbitrary large
accuracy. The equilibrium distance $r_0$ is then determined by the
condition that the total energy is at a minimum:
\[ W(r_0) = \frac{e^2}{r_0} + E(r_0) \]
The curve $W(r)$ has a behavior like that shown in the figure, if
we assume that zero energy corresponds to the molecule which is
dissociated into a neutral hydrogen atom and a ionized atom at an
infinite distance. The equilibrium distance
\begin{center}
%\begin{figure}[t]
\epsfysize=3.5cm \epsfxsize=6cm %
\centerline{\epsffile{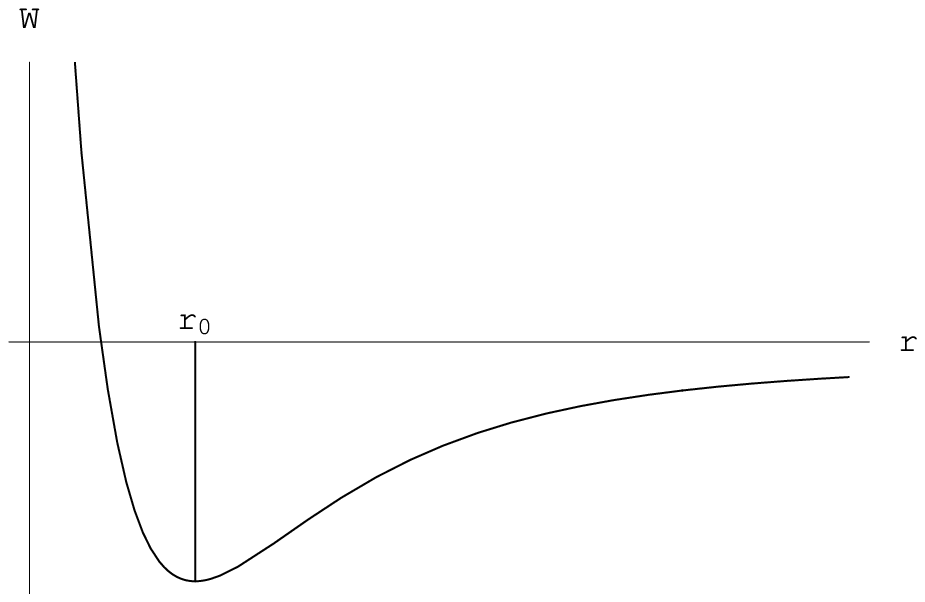}}
%\caption{}
%\label{app2fig}
%\end{figure}
\end{center} \vspace{-0.5truecm}
has been theoretically evaluated by Burrau\footnote{Majorana
refers here to the paper by \O. Burrau, {\it Berechnung des
Energiewertes des Wasserstoffmolekel-Ions (H2+) im Normalzustand},
Kgl. Danske Videnskab. Selskab, Mat. Fys. Medd. {\bf 7} (1927)
14.} finding $r_0 = 1.05 \cdot 10 ^{-8}$ cm and, for the
corresponding energy, $W(r_0) = - 2.75$ electron-volt. Both these
results have been fully confirmed by observations on the spectrum
emitted by the neutral or ionized molecule, that indirectly
depends on them.  \\
What is the origin of the force $\dps F = + \frac{\drm W}{\drm r}$
that tends to get close the two hydrogen nuclei when they are at a
distance larger than $r_0$? The answer given by Quantum Mechanics
to this question is surprising since it seems to show that, beside
certain polarization forces which can be foreseen by Classical
Mechanics, a predominant role is played by a completely novel kind
of forces, the so-called \ul{resonance forces}.  \\
Let us suppose the distance $r$ to be large with respect to the
radius of the neutral hydrogen atom ($\sim 0.5 \cdot 10^{-8} cm$).
Then the electron undergoes the action of one or the other of the
two protons, around each of them can classically describe closed
orbits. The system composed by the electron and the nucleus around
which it orbits forms a neutral hydrogen atom, so that our
molecule results to be essentially composed by one neutral atom
and one proton at a certain distance from the first. The neutral
hydrogen atom in its ground state has a charge distribution with
spherical symmetry, classically meaning that all the orientations
of the electronic orbit are equally possible, and the negative
charge density exponentially decreases with the distance in such a
way that the atomic radius can practically be considered as
finite; it follows that no electric field is generated outside a
neutral hydrogen atom, and thus no action can be exerted  on a
proton at a distance $r$ which is large compared to the atomic
dimensions. However the neutral atom can be polarized under the
action of the external proton and acquire an electric moment along
the proton-neutral atom direction, and from the interaction of
this electric moment with the non uniform field generated by the
proton it comes out an attractive force which tends to combine the
atom and the ion in a molecular system. \\
The \ul{polarization forces}, which can be easily predicted with
classical arguments, can give origin alone to molecular compounds,
that however are characterized by a pronounced fleetingness. More
stable compounds can only be obtained if other forces are
considered in addition to the polarization ones. In the polar
molecules, composed of two ions of different sign charge, such
forces are essentially given by the electrostatic attraction
between the ions; for example the $H Cl$ molecule is kept together
essentially by the mutual attraction between the $H^+$ positive
ion and the $Cl^-$ negative one. However in molecule composed by
two neutral atoms, or by a neutral atom and a ionized one, as in
the case of the molecular ion $H_2^+$, the \ul{chemical affinity}
is essentially driven by the phenomenon of the resonance,
according to the meaning assumed by this word in the novel
Mechanics, which has no parallel in Classical Mechanics.\\
When we study, from the Quantum Mechanics point of view, the
motion of the electron in the field of the two protons, assumed to
be fixed at a very large mutual distance $r$, at a first
approximation we can determine the energy levels by assuming that
the electron  should move around the proton $A$ (or $B$) and
neglecting the influence of the other proton in $B$ (or $A$),
which exerts a weak perturbative action due to its distance. For
the lowest energy eigenvalue $E_0$ we thus obtain a state $S$
corresponding to the formation of a neutral atom in its ground
state consisting of the electron and the nucleus $A$, and a state
$S'$ corresponding to a neutral atom composed by the electron and
the nucleus $B$. Now if we take into account the perturbation that
in both cases is exerted on the neutral atom by the positive ion,
we again find, as long as the perturbation is small, not two
eigenvalues equal to $E_0$ but two eigenvalues $E_1$ and $E_2$
which are slightly different from $E_0$ and both close to this
value; however the quantum states corresponding to them, let be
$T_1$ and $T_2$, are not separately close to $S$ e $S'$, since,
due to the fact that the potential field where the electron moves
is symmetric with respect to the middle point of $AB$, the same
symmetry must be shown, for what said above, by the effective
states $T_1$ and $T_2$ of the electron, while it is not separately
shown by $S$ and $S'$.\\
According to the model representation of the quantum states
introduced above, $S$ consists of a set of electronic orbits
around $A$, and analogously $S'$ of a set of orbits around $B$,
while the true quantum states of the system $T_1$ and $T_2$ each
correspond, at a first approximation for very large $r$, for one
half to the orbits in $S$ and for the other half to those in $S'$.
The computations prove, that for sufficiently large nuclear
distances, the mean value of the perturbed eigenvalue $E_1$ and
$E_2$ coincides closely to the single unperturbed value $E_0$,
while their difference is not negligible and has a conclusive
importance in the present as well as in infinite other analogous
cases of the study of the chemical reactions. We can thus suppose
that $E_1 < E_0$ but $E_2 > E_0$, and then $T_1$ will be the
ground state of the electron, while $T_2$ will correspond to the
excited state with a slightly higher energy. \\
The electron in the $T_1$ state, as well as in the $T_2$ state,
spends half of its time around the nucleus $A$ and the other half
around the nucleus $B$. We can also estimate the mean frequency of
the periodic transit of the electron from $A$ to $B$ and
viceversa, or of the neutral or ionized state exchange between the
two atoms, thus finding
\[ \n = \frac{E_2 - E_1}{h} \]
where $h$ is the Planck constant. For large values of $r$, $E_2 -
E_1$ decreases according to an exponential-like curve and thus the
exchange frequency rapidly tends to zero, this meaning that the
electron which was initially placed around $A$ remains here for an
increasingly larger time, as expected from a classical point of
view. \\
If the electron is in the state $T_1$, that is in its ground
state, its energy ($E_1$) is lower than he would have without the
mentioned exchange effect between nuclei $A$ and $B$. This
occurrence gives origin to a novel kind of attractive forces among
the nuclei, in addition to the polarization forces considered
above, and are exactly the dominant cause of the molecular
bonding. \\
The resonance forces, as said, has no analogy in Classical
Mechanics. However, as long as the analogy leading to the
correspondence between a quantum state and a statistical set of
classical motions can hold, the two states $T_1$ and $T_2$, where
the resonance forces have opposite sign too, each one result
composed identically by half of both the original unperturbed
states $S$ and $S'$. This, however, is true only at a certain
approximation, that is exactly at the approximation where we can
neglect the resonance forces. For an exact computation taking into
account the resonance forces we must necessarily use Quantum
Mechanics, and thus find a \ul{qualitative} difference in the
structure of the two quantum states that manifest itself mainly in
the intermediate region between $A$ and $B$ through which a
periodic transit of the electron between one atom and the other
takes place, according to a mechanism that cannot be described by
Classical Mechanics. Such a qualitative difference is purely
formal in nature and we can deal with it only by introducing the
wave function $\psi(x,y,z)$ that, as known, corresponds in Quantum
Mechanics to any possible state of the electron. The modulus of
the square of $\psi$, which can also be a complex quantity, gives
the probability that the electron lies in the volume unit around a
generic point $x,y,z$. The wave function $\psi$ must then satisfy
to a linear differential equation and thus we can always multiply
$\psi$ in any point by a fixed real or complex number of modulus
$1$, this constraint being required by the normalization condition
\[ \int \! | \psi^2 | \, \drm x \, \drm y \, \drm z = 1 \]
which is necessary for the mentioned physical interpretation of
$|\psi^2|$. The multiplication of $\psi$ by a constant of modulus
$1$ leaves unaltered the spatial distribution of the electronic
charge, and has in general no physical meaning. Now we will
formally define the reflection of a quantum state with respect to
the middle point $O$ between the two nuclei $A$ and $B$ directly
on the wave function $\psi$, by setting
\[ \psi(x,y,z) = \psi'(-x,-y,-z) \]
in a coordinate frame with origin in $O$. If $\psi$ should
represent a symmetric quantum states, and thus invariant by
reflection in $O$, the reflected wave function  $\psi'$ must have
the same physical meaning of $\psi$ and thus differ from $\psi$,
for what said, for a real or complex constant factor of modulus
$1$. Moreover such a constant factor has to be $\pm 1$, since its
square must give the unity, due to the fact that by a further
reflection of $\psi'$ with respect to the point $O$ we again
obtain the initial wave function $\psi$. \\
For all the states of the system we than must have:
\[ \psi (x,y,z) = \pm \psi (-x,-y,-z) \]
where the $+$ sign holds for a part of them, and the $-$ one for
the others. The formal difference between the $T_1$ and $T_2$
states considered above consists precisely in the fact that, in
the previous equation, the upper sign holds for $T_1$ while the
lower one for $T_2$. The symmetry with respect to one point and,
in general, any symmetry property, determines a formal splitting
of the state of the system in two or more sectors, an important
property of this splitting being that no transition between
different sectors can be induced by external perturbations
respecting the symmetries shown by the constraints of the system.
Thus in systems containing two electrons, we have two kinds of not
combinable states which are determined by the fact that the wave
function, which now depends on the coordinates of both the
electrons, remains unaltered or changes its sign by exchanging the
two identical particles. In the special case of the helium atom
this gives rise to the well known spectroscopic appearance of two
distinct elements: parahelium and {\it orthohelium}. \\
The theory of the chemical affinity between the neutral hydrogen
atom and the ionized one, which we have considered until now, can
be extended to the study of the neutral hydrogen molecule and,
more in general, of all the molecules resulting from two equal
neutral atoms. Instead of only one electron moving around two
fixed protons, for the neutral hydrogen molecule we should
consider \ul{two} electrons moving in the same field, neglecting
at a first approximation their mutual repulsion. The stability of
the molecule can then be understood by assuming that each of the
two electrons lies in the $T_1$ state, corresponding to attractive
resonance forces. According to F. Hund we can say that the
hydrogen molecule  is kept together by two ``bounding'' electrons.
However the interaction between the two electrons is so large to
leave only a qualitative explanation for the schematic theory by
Hund, but in principle we could predict exactly all the properties
of the hydrogen molecule, by solving with a sufficient precision
the equations introduced by Quantum Mechanics. In this way, with
appropriate mathematical methods, we can effectively determine the
chemical affinity between two neutral hydrogen atoms with only
theoretical considerations, and the theoretical value agrees with
the experimental one, given the precision of the computation
imposed by practical reasons. \\
For molecules different from the hydrogen one, the theory of the
chemical affinity is considerably more complex, due both to the
larger number of electrons to be considered and to the Pauli
principle, forbidding the simultaneous presence of more than two
electrons in the same state; however the different theories of the
chemical affinity proposed in the last years, each of which has an
applicability range more or less large, practically consist in the
search for approximated computation methods for a mathematical
problem that is exactly determined in itself, and not in the
enunciation of novel physical principles. \\
Then it is possible to bring the theory of the valence saturations
back to more general principles of Physics. Quantum Mechanics
opens the road to the logic unification of all the sciences having
the inorganic world as a common object of study.

\section*{Acknowledgments}

\noindent I am indebted with Dr. Alberto De Gregorio and Prof.
Erasmo Recami for fruitful discussions and suggestion about the
present work.

\vspace{1truecm}

% ----------------------------------------------------------------

\end{document}